\documentclass[floatfix,aps,prl,twocolumn,  superscriptaddress]{revtex4-2}

\usepackage{upgreek}
\usepackage{graphicx}
\usepackage[]{epsfig}
\usepackage{times,amsmath,amssymb}
\usepackage{textcomp}
\usepackage{color}
\usepackage[colorlinks = true,
            linkcolor = blue,
            urlcolor  = blue,
            citecolor = blue,
            anchorcolor = blue]{hyperref}

\begin{document}

\title{RFSoC-based radio-frequency reflectometry in gate-defined bilayer graphene quantum devices}

\author{Motoya Shinozaki}
\email[]{motoya.shinozaki.c1@tohoku.ac.jp}
\affiliation{WPI Advanced Institute for Materials Research, Tohoku University, 2-1-1 Katahira, Aoba-ku, Sendai 980-8577, Japan}

\author{Tomoya Johmen}
\affiliation{Research Institute of Electrical Communication, Tohoku University, 2-1-1 Katahira, Aoba-ku, Sendai 980-8577, Japan}
\affiliation{Department of Electronic Engineering, Graduate School of Engineering, Tohoku University, Aoba 6-6-05, Aramaki, Aoba-Ku, Sendai 980-8579, Japan}

\author{Aruto Hosaka}
\affiliation{Information Technology R\&D Center, Mitsubishi Electric Corporation, Kamakura 247-8501, Japan}

\author{Takumi Seo}
\affiliation{Research Institute of Electrical Communication, Tohoku University, 2-1-1 Katahira, Aoba-ku, Sendai 980-8577, Japan}
\affiliation{Department of Electronic Engineering, Graduate School of Engineering, Tohoku University, Aoba 6-6-05, Aramaki, Aoba-Ku, Sendai 980-8579, Japan}

\author{Shunsuke Yashima}
\affiliation{Research Institute of Electrical Communication, Tohoku University, 2-1-1 Katahira, Aoba-ku, Sendai 980-8577, Japan}
\affiliation{Department of Electronic Engineering, Graduate School of Engineering, Tohoku University, Aoba 6-6-05, Aramaki, Aoba-Ku, Sendai 980-8579, Japan}

\author{Akitomi Shirachi}
\affiliation{Research Institute of Electrical Communication, Tohoku University, 2-1-1 Katahira, Aoba-ku, Sendai 980-8577, Japan}
\affiliation{Department of Electronic Engineering, Graduate School of Engineering, Tohoku University, Aoba 6-6-05, Aramaki, Aoba-Ku, Sendai 980-8579, Japan}

\author{Kosuke Noro}
\affiliation{Research Institute of Electrical Communication, Tohoku University, 2-1-1 Katahira, Aoba-ku, Sendai 980-8577, Japan}
\affiliation{Department of Electronic Engineering, Graduate School of Engineering, Tohoku University, Aoba 6-6-05, Aramaki, Aoba-Ku, Sendai 980-8579, Japan}

\author{Shoichi Sato}
\affiliation{Research Institute of Electrical Communication, Tohoku University, 2-1-1 Katahira, Aoba-ku, Sendai 980-8577, Japan}

\author{Takashi Kumasaka}
\affiliation{Research Institute of Electrical Communication, Tohoku University, 2-1-1 Katahira, Aoba-ku, Sendai 980-8577, Japan}

\author{Tsuyoshi Yoshida}
\affiliation{Information Technology R\&D Center, Mitsubishi Electric Corporation, Kamakura 247-8501, Japan}

\author{Tomohiro Otsuka}
\email[]{tomohiro.otsuka@tohoku.ac.jp}
\affiliation{WPI Advanced Institute for Materials Research, Tohoku University, 2-1-1 Katahira, Aoba-ku, Sendai 980-8577, Japan}
\affiliation{Research Institute of Electrical Communication, Tohoku University, 2-1-1 Katahira, Aoba-ku, Sendai 980-8577, Japan}
\affiliation{Department of Electronic Engineering, Graduate School of Engineering, Tohoku University, Aoba 6-6-05, Aramaki, Aoba-Ku, Sendai 980-8579, Japan}
\affiliation{Center for Science and Innovation in Spintronics, Tohoku University, 2-1-1 Katahira, Aoba-ku, Sendai 980-8577, Japan}
\affiliation{Center for Emergent Matter Science, RIKEN, 2-1 Hirosawa, Wako, Saitama 351-0198, Japan}

\date{\today}

\begin{abstract}
Quantum computers require both scalability and high performance for practical applications. 
While semiconductor quantum dots are promising candidates for quantum bits, the complexity of measurement setups poses an important challenge for scaling up these devices. 
Here, radio-frequency system-on-chip (RFSoC) technology is exepcted for a promising approach that combines scalability with flexibility. 
In this paper, we demonstrate RF reflectometry in gate-defined bilayer graphene quantum devices using RFSoC-based measurement architecture. 
By controlling the confinement strength through gate voltages, we achieve both Fabry-Pérot interferometer and quantum dot operations in a single device.
Although impedance matching conditions currently limit the measurement sensitivity, we identify pathways for optimization through tunnel barrier engineering and resonator design. 
These results represent a step toward integrating high-bandwidth measurements with scalable quantum devices.
\end{abstract}

\maketitle
Quantum computers have emerged as a promising technology with the potential to revolutionize various fields. 
Several physical systems for quantum bits have been proposed, including superconducting circuits, trapped ions, and nitrogen-vacancy centers in diamonds~\cite{popkin2016quest}. 
Among these approaches, semiconductor quantum dots~\cite{tarucha1996shell, kouwenhoven1997excitation, kouwenhoven2001few} have attracted attention owing to their potential compatibility with existing semiconductor manufacturing processes~\cite{Elsayed2024}. 
Advances in quantum state control and readout have been achieved through increasingly complicated measurement systems, making it challenging to scale beyond a few multiple quantum dots.
The vision of quantum computers with millions of quantum bits has been widely discussed, yet the increasing complexity of measurement systems may become a fundamental bottleneck~\cite{Bharti2022, malinowski2023wire}.
Here, radio-frequency system-on-chip (RFSoC) technology, originally developed for next-generation communication systems, has been utilized to address these issues~\cite{Leamdro_Qick2022, Park_icarus2022, Ding2024exp}.

To realize practical quantum computers, both scalability and performance must be addressed simultaneously. 
Various approaches have been explored to improve performance, one of which is the development of novel materials platforms. 
So far conventional semiconductors such as GaAs~\cite{koppens2006driven, yoneda2014fast}, Si~\cite{muhonen2014storing, Veldhorst2014, 2018YonedaNatureNano}, and ZnO~\cite{Noro2024} have been studied for quantum devices, bilayer graphene (BLG)~\cite{zhang2009direct} has emerged as a promising candidate, attracting attention not only for its unique electronic properties but also for quantum applications.
For example, not only the spin-degree of freedom, valley-degree-of-freedom-based quantum states have highlighted those long relaxation times and potential coherence~\cite{banszerus2022spin, garreis2024long}, driving continued research into graphene-based quantum devices.
The characterization and control of these quantum systems require advanced measurement techniques, among which radio-frequency (RF) reflectometry has been established as an important technique to observe quantum dynamics with broad bandwidth~\cite{qin2006radio, reilly2007fast, barthel2009rapid}. 
While this technique has been demonstrated in GaAs and Si systems, its demonstration in graphene devices has been limited with a few reported cases~\cite{banszerus2021dispersive, johmen2023radio, Reuckriegel_electric2024}.
This is because of the technical expertise of combining high-quality graphene device fabrication with RF measurement techniques.
These challenges motivate the development of measurement architectures that can combine scalability with versatility.

In this study, we apply RFSoC techniques to semiconductor quantum systems and demonstrate RF reflectometry in gate-defined bilayer graphene quantum devices utilizing quantum instrumentation control kit (QICK). 
Our results advance both the scalability and performance aspects of quantum device measurements.

\begin{figure}
\begin{center}
  \includegraphics{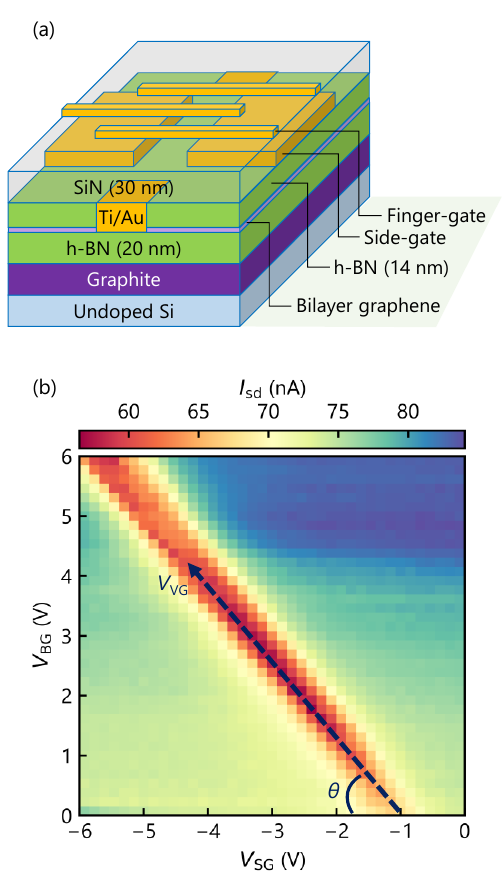}
  \caption{(a) Schematic of the layer structure of the device.
  (b) Back-gate voltage $V_\mathrm{BG}$ and side-gate voltage $V_\mathrm{SG}$ dependence on the source-drain current $I_\mathrm{sd}$. 
  }
  \label{fig1}
\end{center}
\end{figure}

Figure~\ref{fig1}(a) illustrates the layer structure of the BLG device. 
Both graphene and hexagonal boron nitride (hBN) layers are mechanically exfoliated using the scotch-tape method. 
These layers are stacked on an undoped-silicon substrate through a pick-and-release process using polypropylene carbonate coated polydimethylsiloxane stamp~\cite{pizzocchero2016hot, Iwasaki2020}.
The thickness of top and bottom hBN layers are 14 and 20~nm, respectiverely.
Ti/Au electrodes serving as source-drain, side-gate, and finger-gate contacts are fabricated by electron beam evaporation. 
A silicon nitride layer of 30~nm is deposited as a top insulator by chemical vapor deposition. 
The bottom graphite layer functions as a back-gate, which combined with the insulating substrate reduces stray capacitance and enables us to apply RF signal to the device~\cite{johmen2023radio}. 
The finger-gates are designed with a width of approximately 40~nm and aligned with 40~nm gaps. 
We operate only one finger-gate in this study.

Figure~\ref{fig1}(b) shows the back-gate voltage $V_\mathrm{BG}$ and side-gate voltage $V_\mathrm{SG}$ dependence on the source-drain current $I_\mathrm{sd}$, with the finger-gate voltage $V_\mathrm{FG}$ fixed at 0~V. 
All measurements are conducted at a temperature of 2.3~K using a helium decompression refrigerator.
A region of suppressed conductance appears along the diagonal direction in the figure.
This reflects the band gap opening in BLG induced by the vertical electric field, proportional to $|V_\mathrm{BG} - V_\mathrm{SG}|$~\cite{zhang2009direct}.
Because of this band gap opening, a one-dimensional channel can be defined in the gap of side-gate electrodes.

\begin{figure}
\begin{center}
  \includegraphics{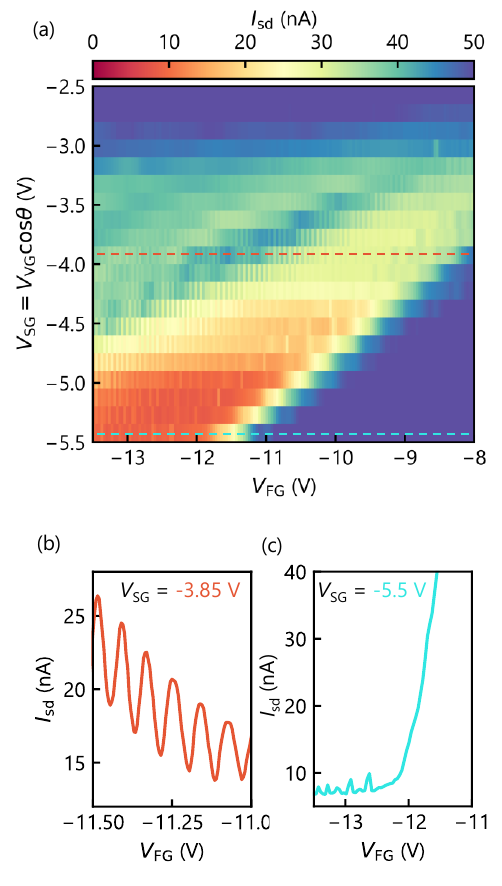}
  \caption{(a) The dependence of $I_\mathrm{sd}$ on $V_\mathrm{FG}$ and the virtual gate voltage $V_\mathrm{VG}$. 
  (b) An extracted $I_\mathrm{sd}$ at $V_\mathrm{SG}=-3.85$~V and (c) $V_\mathrm{SG}=-5.5$~V.
  }
  \label{fig2}
\end{center}
\end{figure}

\begin{figure*}
\begin{center}
  \includegraphics{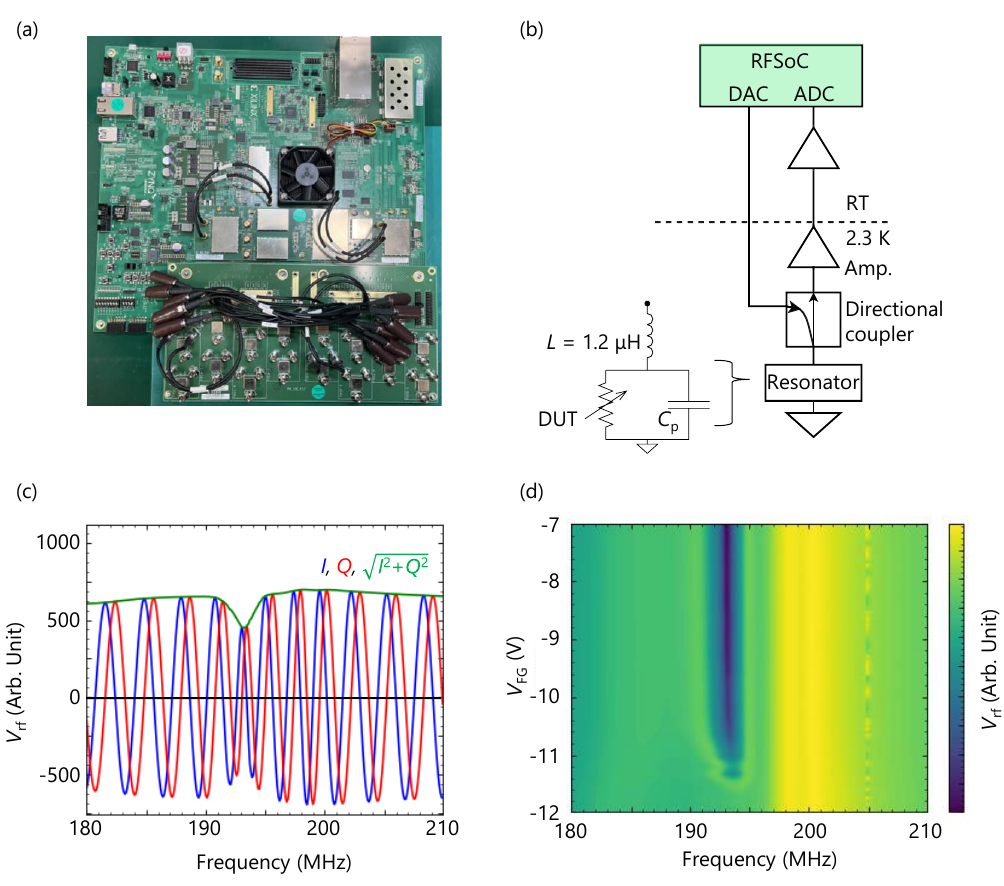}
  \caption{(a) The Radio-frequency system-on-chip (RFSoC) measurement borad. 
  (b) A measurement setup for radio-frequency reflectometry with RFSoC.
  (c) A frequency dependence of the demodulated signals at $V_\mathrm{FG}=-10$~V.
  (d) The $V_\mathrm{FG}$ dependence of the resonator characteristics.
  }
  \label{fig3}
\end{center}
\end{figure*}

In order to form a quantum dot, we apply $V_\mathrm{FG}$ to the gate-defined one-dimensional channel. 
Figure~\ref{fig2}(a) shows the dependence of $I_\mathrm{sd}$ on $V_\mathrm{FG}$ and the virtual gate voltage $V_\mathrm{VG}$. Here, $V_\mathrm{VG}$ is defined by considering the electrostatic coupling between the side- and back-gates to the BLG layer, as illustrated in Fig.~\ref{fig1}(b), and is represented by its $V_\mathrm{SG}$ component. 
Note that $V_\mathrm{BG}$ is swept along with $V_\mathrm{SG}$ to maintain their relationship.
With increasing $V_\mathrm{SG}$, oscillation-like features appear around $V_\mathrm{SG}=-3.85$~V. 
A typical trace is shown in Fig.~\ref{fig2}(b). 
These oscillations indicate the formation of a Fabry–Pérot (FP) interferometer in our device. 
The combination of positive $V_\mathrm{BG}$ and negative $V_\mathrm{FG}$ defines a n-p-n cavity in the channel, resulting in FP interference~\cite{young2009quantum, masubuchi2013fabrication, ahmad2019fabry, Portolés_2022}. 
As we further increase $V_\mathrm{VG}$, $I_\mathrm{sd}$ becomes clearly suppressed.
In this region, the larger band gap opening leads to a narrower channel, enabling stronger electron confinement by $V_\mathrm{FG}$. 
The strong confinement allows us to electrically define a quantum dot directly beneath the finger-gate electrode. 
Several peaks are observed in the pinch-off region at $V_\mathrm{SG}=-5.5$~V, as shown in Fig.~\ref{fig2}(c), indicating Coulomb oscillations from the quantum dot.
Thus, we demonstrate a crossover from FP interference to quantum dot behavior in a single device by tuning the confinement strength.

Next, we develop the RF measurement setup.
The complexity of RF measurement setups poses an obstacle to scaling up quantum systems.
Various approaches toward large-scale control have been investigated, including frequency-multiplexed readout to reduce the number of control lines~\cite{al2013cryogenic} and on-chip control circuits integrated with quantum devices~\cite{schaal2019cmos, Xu_onchip2020, pauka2021cryogenic}.
The RFSoC technology presents a promising approach that combines scalability with flexibility~\cite{Leamdro_Qick2022, Park_icarus2022, Maetani_2024, Ding2024exp}. 
Figure~\ref{fig3}(a) shows the RFSoC measurement board Zynq\textregistered~UltraScale+\texttrademark~RFSoC ZCU216 evaluation kit~\cite{ZCU216}.
The RFSoC-based platform can be controlled using a Python interface provided by open-source software called the QICK~\cite{QICK_git, Leamdro_Qick2022}.
By integrating digital-to-analog converters (DACs), analog-to-digital converters (ADCs), and field-programmable gate arrays (FPGAs) on a single board, RFSoC enables RF signal processing capabilities with programmability and adaptability to various quantum devices. 
The on-board FPGA also provides real-time signal processing and feedback control, which has become increasingly important for quantum measurements~\cite{nakajima2021real, fujiwara2023wide}. 
Several studies have explored RFSoC's potential applications in quantum devices. 
While preliminary experiments suggest its feasibility, the RFSoC-based RF reflectometry in semiconductor quantum dots remains to be demonstrated. 
This technology offers a promising platform to bridge the gap between traditional measurement systems and future large-scale quantum computers.

Figure~\ref{fig3}(b) shows the measurement setup for RF reflectometry. 
An RF signal is applied to a resonator consisting of the graphene device, chip inductor, and capacitors from both the printed circuit board (PCB) and the device. 
The reflected signal from the resonator is amplified and then received by the ADC. 
The received signal is digitized and demodulated into In-phase ($I$) and Quadrature-phase ($Q$) components using the built-in local oscillator in the RFSoC.
Figure~\ref{fig3}(c) shows a typical resonator response obtained at $V_\mathrm{FG}=-10$~V.
The amplitude $\sqrt{I^2+Q^2}$ exhibits a clear resonance around 193~MHz, demonstrating the capability of RFSoC to perform RF reflectometry. 
The $V_\mathrm{FG}$ dependence of the resonator characteristics is shown in Fig.~\ref{fig3}(d).
Here, we fix $V_\mathrm{BG}$ and $V_\mathrm{SG}$ at 5.5 and -5.4~V, respectively. 
The resonance is modulated by varying $V_\mathrm{FG}$. 
Considering an inductance $L$ of 1.2~$\upmu$H, the total capacitance $C_\mathrm{p}$ in the resonator is calculated to be 0.6~pF from the resonance frequency, which primarily originates from the PCB's stray capacitance.
The stray capacitance from the device itself can be neglected owing to our device structure employing a microscale graphite back-gate~\cite{johmen2023radio}.

\begin{figure}
\begin{center}
  \includegraphics{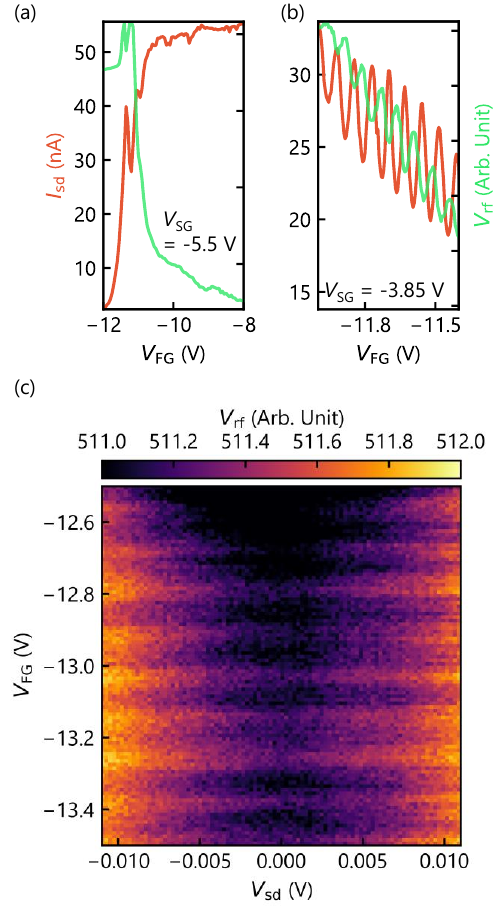}
  \caption{$V_\mathrm{rf}$ and $I_\mathrm{sd}$ with sweeping $V_\mathrm{FG}$ for both (a) strong and (b) weak confinement states.
  (c) Coulomb diamonds probed by $V_\mathrm{rf}$.
  }
  \label{fig4}
\end{center}
\end{figure}

We compare $V_\mathrm{rf}$ at 193~MHz and $I_\mathrm{sd}$ by sweeping $V_\mathrm{FG}$ for both strong and weak confinement states corresponding to Figs.~\ref{fig2}(b) and (c), as shown in Figs.~\ref{fig4}(a) and (b). 
The $V_\mathrm{rf}$ trace clearly reflects the changes in $I_\mathrm{sd}$, demonstrating that $V_\mathrm{rf}$ can probe the device conductance.
Furthermore, we measure $V_\mathrm{rf}$ as a function of $V_\mathrm{sd}$ and $V_\mathrm{FG}$, as shown in Fig.~\ref{fig4}. 
Periodic suppressions in $V_\mathrm{rf}$ appear along $V_\mathrm{FG}$, which is Coulomb diamonds indicating quantum dot formation. 
Using the constant interaction model, we extract a finger-gate capacitance of approximately 4~aF from the smallest diamond shape. 
This value agrees well with previous studies of gate-defined graphene quantum dots~\cite{banszerus2018gate, Eich2018spin, banszerus2021dispersive}.

Here, the RF reflectometry has potential for improved sensitivity. 
To achieve more sensitive readout, impedance matching is important for optimal RF reflectometry response and is satisfied at a device conductance of around 24~$\upmu$S in our resonator.
This condition corresponds to $I_\mathrm{sd}=120$~nA, which is far from our measurement results. 
Consequently, our RF reflectometry operates at a point of poor sensitivity.
Previous studies have also shown Coulomb peaks below 10~$\upmu$S, indicating the challenge in demonstrating resistive readout with RF reflectometry. 
To address this issue, we propose two approaches. 
The first is to increase the device conductance. 
The low conductance originates from tunnel barriers between the dot and source/drain electrodes, rather than contact resistivity, as the latter is negligible in graphene devices due to being much smaller than the quantum resistance.
Additional gate electrodes to tune the tunnel barrier can modulate device conductance~\cite{Reuckriegel_electric2024, Tong2024pauli}, resulting in improvement measurement sensitivity.
The second approach is to develop a resonator with a variable matching circuit. 
By connecting additional capacitors in parallel to the resonator, the impedance matching condition can be satisfied at lower conductance values~\cite{ares2016sensitive, Eggli2023cryo, apostolidis2024quantum}.

In conclusion, we demonstrate RFSoC-based radio-frequency reflectometry measurements in gate-defined bilayer graphene quantum devices.
By controlling the confinement strength through gate voltages, we define both Fabry-Pérot interferometer and quantum dot operation in a single device. 
The RF reflectometry measurement reveals clear Coulomb diamonds, from which we extract a finger-gate capacitance of approximately 4~aF. 
While the current implementation operates at poor sensitivity due to impedance mismatching, we propose promising approaches to enhance the measurement sensitivity. 
Our results show the potential of RFSoC-based architecture for scalable quantum measurements, contributing to the development of large-scale quantum devices.

\section{Acknowledgements}
The authors thank S. Iwakiri, 
RIEC Fundamental Technology Center and the Laboratory for Nanoelectronics and Spintronics 
for fruitful discussions and technical supports.
Part of this work is supported by 
Grants-in-Aid for Scientific Research (21K18592, 23H01789, 23H04490),
CREST (JPMJCR23A2), JST, 
and FRiD Tohoku University.

\section{AUTHOR DECLARATIONS}
\subsection{Conflict of Interest}
The authors have no conflicts to disclose.
\subsection{Author Contributions}
\textbf{Motoya Shinozaki:} Conceptualization (lead); Data Curation (lead); Investigation (lead); Methodology (lead); Resources (equal);  Software (equal); Visualization (lead); Writing/Original Draft (lead); Writing/Review \& Editing (equal). 
\textbf{Tomoya Johmen:} Conceptualization (equal); Data Curation (equal); Investigation (equal); Methodology (equal); Resources (lead); Writing/Review \& Editing (equal);
\textbf{Aruto Hosaka:} Conceptualization (equal); Investigation (equal); Methodology (equal); Resources (equal);  Software (lead); Visualization (equal); Writing/Review \& Editing (equal). 
\textbf{Takumi Seo:} Resources (equal); Writing/Review \& Editing (equal). 
\textbf{Shunsuke Yashima:} Resources (equal); Visualization (equal); Writing/Review \& Editing (equal). 
\textbf{Akitomi Shirachi:} Resources (equal); Writing/Review \& Editing (equal). 
\textbf{Kosuke Noro:} Resources (equal); Writing/Review \& Editing (equal). 
\textbf{Shoichi Sato:} Resources (equal); Writing/Review \& Editing (equal);
\textbf{Takeshi Kumasaka:} Resources (equal); Writing/Review \& Editing (equal);
\textbf{Tsuyoshi Yoshida:} Conceptualization (equal); Resources (equal); Funding Acquisition (equal); Writing/Review \& Editing (equal). 
\textbf{Tomohiro Otsuka:} Conceptualization (equal); Methodology (equal); Funding Acquisition (lead); Supervision (lead); Writing/Review \& Editing (lead).

\section{DATA AVAILABILITY}
The data that support the findings of this study are available from the corresponding authors upon reasonable request.

\bibliography{reference.bib}
\end{document}